\documentclass[a4paper]{jpconf}
\usepackage{graphicx}

\begin{document}
\title{Searching for white dwarf variables in TESS data}

\author{Rhorom Priyatikanto}

\address{Space Science Center, National Research and Innovation Agency, Indonesia}

\ead{rhorom.priyatikanto@brin.go.id}

\begin{abstract}
A sample of 4 thousands Gaia sources with apparent $G$-band magnitudes of $<17$ and trigonometric parallaxes of $>3.33$ milli-arcseconds were cross-matched to the Transiting Exoplanet Survey high level science products and the Full-Frame Images in order to extract the light curves of possible white dwarf variables. Most of the targets in the sample were observed in at least 27-day observing cycle with 30-minutes cadence. Based on Lomb-Scargle periodogram constructed from de-trended light curves, more than 600 sources have indication of periodic variability. This paper presents the early results from the identification.
\end{abstract}

\section{Introduction}
Most of transiting exoplanets were discovered using Kepler space telescope which observes a small patch of starry sky in the constellation of Cygnus \cite{borucki2010}. Following the great success of this mission, NASA is progressing with the second light of Kepler mission known as K2 \cite{howell2014} and also by launching and operating all-sky space-borne survey named Transiting Exoplanet Survey Satellite or TESS in short \cite{ricker2015}. TESS was launched in 2018 and started producing valuable data since July that year.

Orbiting in an elliptical orbit with period of $13.7$ days which is approximately half of lunar month, TESS operates 4 cameras with telescopic lenses. Every camera consists of 4 CCDs with modest sensitivity from 600 to 1000 nm. In total, TESS has $24 \times 90$ square degrees field of view with small portion of blind areas between CCDs. For 27.4 days, TESS observes a particular sector and records data with various cadence. To date, TESS has entered its fourth productive year and acquiring abundant data from 42 sectors covering $\sim90\%$ of the sky. Some sectors overlap such that the sky coverage at some areas may reach a year. Until August 2021, TESS identified 4471 objects of interest. There are at least 154 confirmed exoplanets among them.

The normal operating cycle of TESS starts with capturing the sky with 2-second exposure time which can be stacked into 20-sec, 120-sec, and 1800-sec cadence data. After considering the limited storage posed by the satellite, high cadence data only cover a small portion of the sky around targeted objects. There are approximately $1,000$ small postage stamps (or TESS pixel files, TPF) for astroseismology target stars recorded with 20-sec cadence, $100,000$ postage stamps for priority exoplanet search. The full frame images (FFIs) covering the whole field of view are recorded for every 30 minutes. Based on the acquired FFIs, independent processes can be applied to extract quality data from much more targets of interest. Several groups developed and implemented pipelines to perform such extraction tasks for their specific inquiries. Among others, there are STELLA for stellar flare detection \cite{feinstein2020}, DIAMANTE for detecting exoplanets orbiting FGKM dwarfs and subgiants \cite{montalto2020}, TASOC for astroseismology \cite{lund2017}, TESS-SPOC as the extension of the TESS main products \cite{caldwell2020}, and QLP which produced more than 24-million light curve segments from southern and northern ecliptic hemisphere \cite{huang2020}.

Apart from its major mission for detecting exoplanet and producing target for follow-up observation, data from TESS also triggers discoveries and advanced research in other areas of astronomy. Among others, there are studies about subdwarf blue variables \cite{sahoo2020, baran2021}, up to the dwarf novae in the accreting binary system \cite{court2019}. Inspired by those studies, this current work's objectives are to obtain light curves from TESS data especially for the target stars which are located below main sequence and close to white dwarf. Although TESS observes millions of bright objects, this satellite has its own limitation, mainly related to spatial and spectral resolution. Consequently, this work positions the TESS in the important part of target selection for further follow-up observations. If white dwarf is discussed, then short period binaries and accreting system become interesting issues.

\section{Data and Methods}
To achieve the objectives, a list of target sources was created based on the Gaia Early Data Release 3, just published at the end of 2020 \cite{gaia2016, gaia2021}. Among $2.9$ million sources closer than $300$ pc and brighter than $17$ mag in Gaia $G$ band, objects situated below the main-sequence and bluer than $B_P-R_P=1.5$ were selected. This criteria produced a list containing more than $40,000$ sources while stricter criteria related to the photometric quality was implemented to cut the list leaving approximately 10\% from its initial size. There are 4593 sources were selected as the targets.

Light curves for some of those targets were acquired from the Mikulski Archive for Space Telescopes (MAST) since some high level science products were archived there. TESS-SPOC and QLP produced a large amount of light curves beyond the main TESS products. Both TESS-SPOC and QLP analyzed full frame images and extracted light curves mostly from stars brighter than $13.5$ in $T_{mag}$. The outputs of those two pipelines can be accessed in the MAST programmatically using \textsc{lightkurve} python package \cite{lightkurve2018}. As the complement, \textsc{eleanor} package \cite{feinstein2019} was used to perform independent aperture photometry from target pixel files or TEES cut. This independent analysis enabled the extraction of light curves for dimmer targets. In principle, the output of aperture photometry performed either by TESS-SPOC, QLP, or independently is time series data containing corrected flux and associated error. The flux can be transformed into TESS-magnitude ($T_{mag}$) with baseline from the TESS Input Catalog \cite{stassun2019}.

After acquisition, the next process was light curve median filtering (1-day kernel) to correct the systematic trends or transient variation. Any object with brightening of more than $0.5$ magnitude after filtering is flagged as possible transient. The residual of the filtering was processed to get Lomb-Scargle periodogram as the key for finding periodic signal \cite{vanderplas2018, lomb1976, scargle1982}. The period it self ranges from approximately 35 minutes (close to the data cadence) up to 10 days. Based on the established periodogram, the target is considered as periodic if the maximum power exceeds 3 times the false alarm level. 


\section{Result}
\begin{figure}
    \centering
    \includegraphics[width=\textwidth]{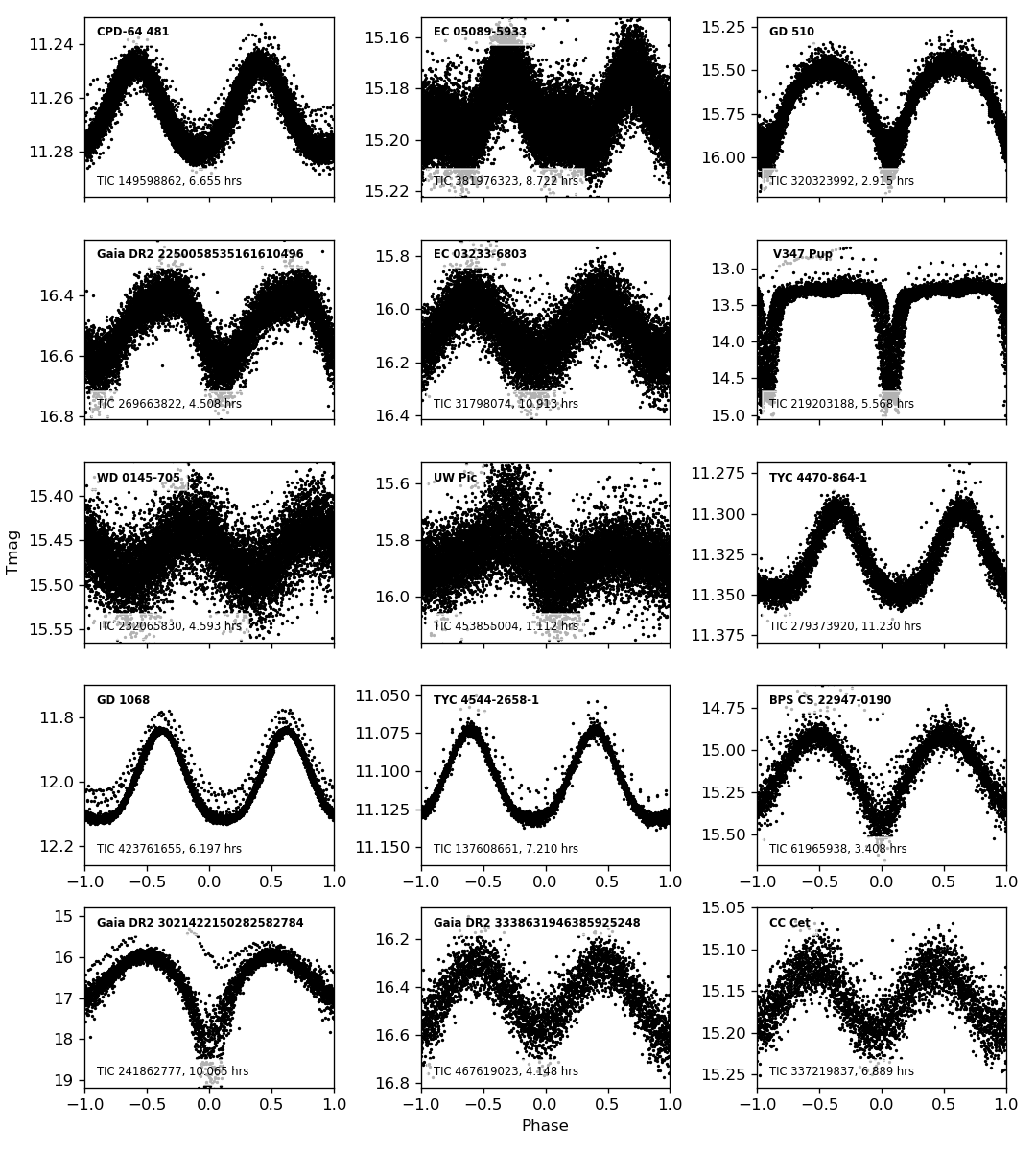}
    \caption{Folded light curves for some frequently observed objects. The name printed on the upper right corner for each panel is the main identification listed in \textsc{Simbad}.}
    \label{most-visited}
\end{figure}

In general, TESS data for two-third of the targets are processed by TESS-SPOC and/or QLP. Light curves from $2,985$ targets were downloaded and analyzed using procedure described before. Independent light curve extraction using \textsc{eleanor} was performed to $3,928$ targets. Interesting objects with clear sign of periodic variability or brightening were flagged. Among 634 of those objects, 382 are periodic with various amplitude (some milli-magnitude to more than one magnitude). There are 46 objects identified having variability period of less than 2 hours. Some well-studied cataclysmic variables were on the list and such object can be used as the benchmark. Figure \ref{most-visited} presents the folded light curves that come from objects observed by TESS a couple of times. From these objects, TESS-SPOC and/or QLP processed the data really well, but flux variation across sectors are obvious for some objects (e.g. GD 1068 in figure \ref{most-visited}). Variable background levels among frames acquired within 27-day orbital period or beyond become the main source of problem to be corrected more properly. Different imaging scheme \cite{montalto2020} or point spread function photometry \cite{nardiello2020} may provide better solution for this problem.

\begin{figure}
    \centering
    \includegraphics[width=\textwidth]{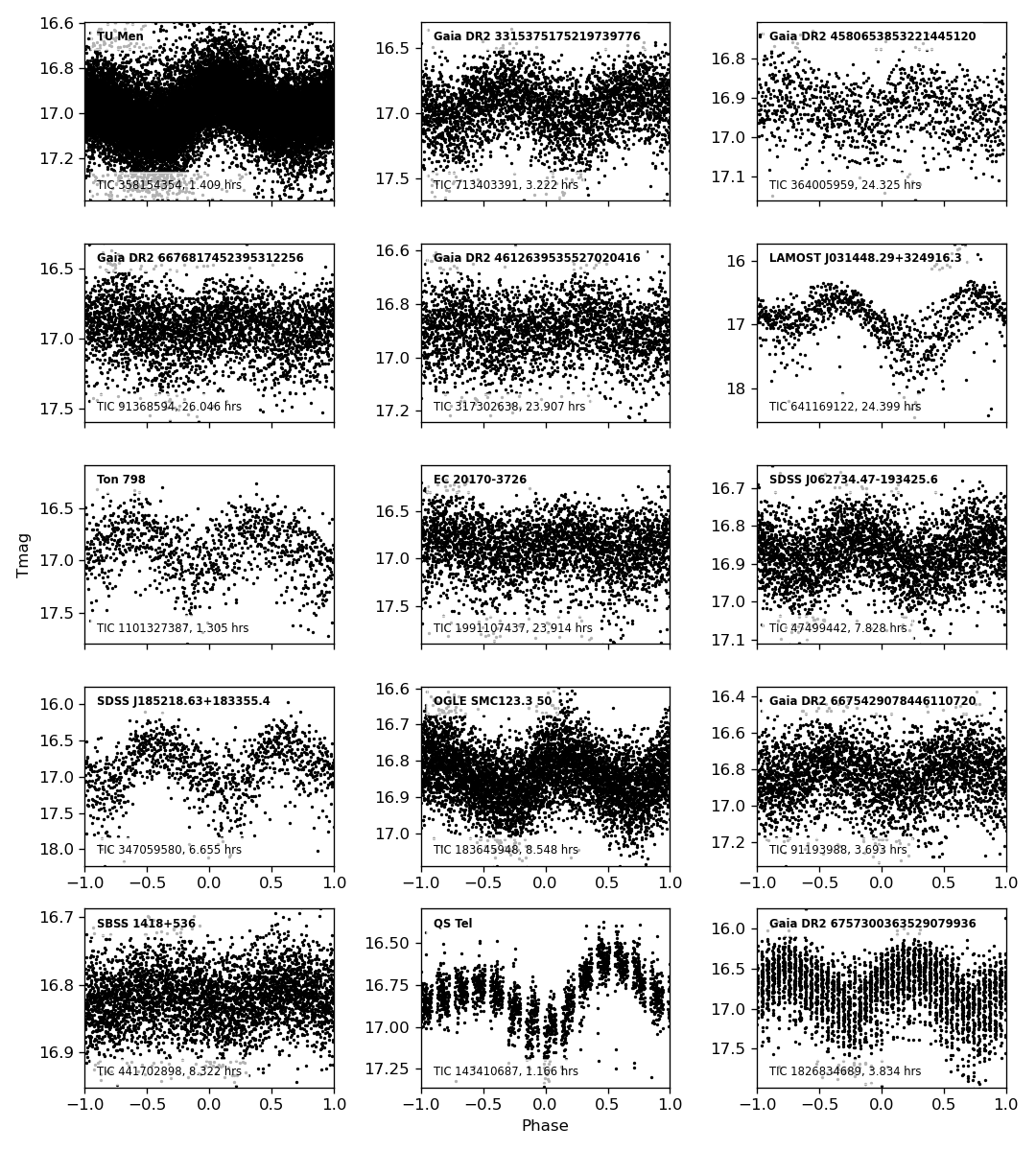}
    \caption{Same as figure \ref{most-visited}, but for some faintest targets.}
    \label{faintest}
\end{figure}

Additionally, light curves from some faintest targets can be seen in figure \ref{faintest}. For the QLP products, Huang \emph{et al}. \cite{huang2020} demonstrated the precision of $10^3$ part per million (ppm) for objects with $T_{mag}\approx13.5$. If the trend is extrapolated, then the expected precision at $T_{mag}\approx17.0$ is at the order of $10^2$ ppt. Crowding field and imperfect data stitching across sectors may decrease the precision of some objects.

\begin{figure}
    \centering
    \includegraphics[width=0.7\textwidth]{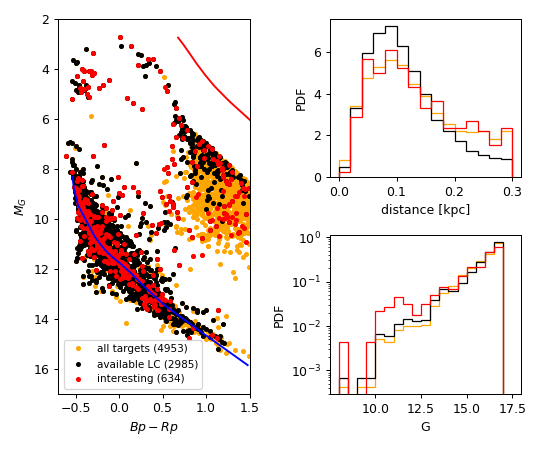}
    \caption{The distribution of $4,593$ targets (black), 2,985 targets with publicly available light curves (orange), and flagged objects (red) in color-magnitude diagram (left), distance space (upper right), and apparent magnitude (lower right).}
    \label{hrdiag}
\end{figure}

The distribution of the targets and interesting/flagged objects is displayed in figure \ref{hrdiag}. Based on the number distribution as a function of $G$, it is obvious that the flagged objects tend to be brighter. On the other hand, a portion of nearby targets are overlooked by TESS-SPOC and QLP as indicated by significantly lower number density at distance of $\sim0.1$ kpc.

The complete table of white dwarf variables needs to be curated and compared to relevant existing catalog, such as the white dwarf catalog \cite{gentile2019, gentile2021}, confirmed cataclysmic variables \cite{abril2020}, or all-sky catalog of UV sources \cite{luciana2017}. Additionally, Baran \emph{et al} \cite{baran2021}. identified 506 blue subdwarf variables from TESS dataset which may coincide with the objects identified in this study.

However, manual screening of the produced light curve was indeed producing list of fascinating objects for further exploration. On top of the list, there is TIC 60040774 which was identified by Gentile Fusillo \emph{et al}. \cite{gentile2019} as probable white dwarf with mass of $0.277$ M$_{\odot}$. It is noteworthy that white dwarf which such low mass is expected to stay in the main-sequence for a period longer than the current age of the universe. Thus, it is likely that this object actually consist of unresolved white dwarf binary with larger mass and observed as a single brighter object. More interestingly, TIC 60040774 shows excess in infrared implying the existence of companion with substellar mass. It can either be planet or brown dwarf, depends on its mass. The system has orbital period of $9.715378$ hours with flux depression of $\sim60\%$ though the actual depression need to be confirmed as the flux from the object is blended with the fluxes from neighboring brighter sources. However, pixel-by-pixel analysis supports the notion that the 9.7-hrs variability comes from a dim blue source previously identified as high proper motion white dwarf. Further photometric and spectroscopic observations are required to solve the mystery about this object. If it is indeed a white dwarf-brown dwarf pair, then this object will be the fourth of its kind \cite{casewell2020}.

Some objects with outburst are indeed observed by TESS. However, most of them are well-studied by astronomers, even monitored by amateur astronomer community. Finding a new understudied object is a challenging process. The other fascinating object is TIC 48448653 with possible outburst observed by TESS around Julian Date $JD = 2458695$. At another time, possible outburst with more than 1 mag brightening was observed by ground-based ASAS-SN telescopes. Apart from that, the periodic variability is unclear.

\section{Closing Remarks}
Even though the analysis of 3,900 objects was performed and the variability of $634$ objects were identified, this is not a finished work. There are more works to do in the future such as cross-matching with other catalog, data mining from databases of multi-wavelength data and machine learning for automatic classification. In near future, this study is expected to generate a list of fascinating objects for follow-up observations to be conducted at astronomical observatories in Indonesia.

\section*{Acknowledgment}
This work is a part of research visit at the University of Southampton hosted by Prof. Christian Knigge and supported by Dr. Simone Scaringi from the Durham University. This work has made use of data from the European Space Agency (ESA) mission {\it Gaia}, processed by the {\it Gaia} Data Processing and Analysis Consortium (DPAC). Funding for the DPAC has been provided by national institutions, in particular the institutions participating in the {\it Gaia} Multilateral Agreement.

\section*{References}
\bibliographystyle{apalike}
\bibliography{refs}
\end{document}